\documentclass[english]{revtex4}
\usepackage[T1]{fontenc}
\usepackage[latin1]{inputenc}
\usepackage{babel}
\usepackage{array}
\usepackage{graphics}
\makeatother
\begin{document}
\title{On the possibility of empirically probing the Bohmian model in terms of the testability of quantum arrival/transit time distribution}
\author{Dipankar Home\footnote{dhome@bosemain.boseinst.ac.in} 
and Alok Kumar Pan\footnote{apan@bosemain.boseinst.ac.in}}
\address{Department of Physics, CAPSS, Bose Institute, Sector-V, Salt Lake, Calcutta
700091, India}
\begin{abstract}
The present article focuses on studying the extent to which the nonuniqueness that is inherent in the standard quantum mechanical calculation of arrival/transit time distribution can be exploited to enable an empirical scrutiny of any causal trajectory model such as the Bohmian scheme. For this purpose, we consider the example of spin-1/2 neutral particles corresponding to a wave packet which passes through a spin rotator(SR) that contains constant magnetic field confined within a region -  in such a case, the transit time distribution can be measured in terms of the spin distribution of particles emerging from the SR. In particular, we investigate the way one can compare the Bohmian predictions obtained for this example with that using one of the quantum approaches, say, the probability current density based scheme. Here the Bohmian calculational procedure involves a couple of critical subtleties that lead to some specific directions for further studies.
\end{abstract}
\maketitle
\section{Introduction and motivation}
Born's interpretation of the squared modulus of a wave function \( ( \left| \psi \right| ^{2})  \) as the probability density of \emph{finding} a particle within a specified region of space is a key ingredient of the standard framework of quantum mechanics, thereby implying that  the standard interpretation of quantum mechanics is inherently \emph{epistemological}. On the other hand, the possibility of an \emph{alternative interpretation} of quantum mechanics by interpreting \( \left| \psi \right| ^{2} \) as the probability density of a particle \emph{being} actually present within a specified region was first suggested by de Broglie\cite{debroglie}.Later, Bohm \cite{bohm}developed the details of such an \emph{ontological} model of quantum mechanics by using the notion of an \emph{observer-independent spacetime trajectory} of an \emph{individual particle} that is determined by its wave function through an equation of motion which is formulated in a way \emph{consistent} with the Schroedinger time evolution. Bohm's model, thus, explicitly refuted the counterarguments (such as the ones put forward by Pauli\cite{pauli}and von Neumann\cite{von}) that claimed to have ruled out the formulation of such a model. Subsequently, much work has been done on various aspects of the Bohmian model\cite{bohm1}. That such a model is \emph{not} unique has  been elaborately discussed\cite{holland2} while different versions of the ontological model of quantum mechanics have been proposed\cite{epstein,roy,holland3,ghirardi}.

Although any such ontological model hinges on the notion of a definite \emph{spacetime track} used to provide a description of the objective
motion of a single particle, such trajectories are \emph{not} directly measurable. Hence these trajectories have been essentially viewed
as conceptual aids for understanding the various features of quantum mechanics. While recently a study has been reported which shows an application of such trajectories as computational aids for solving the time-dependent Schroedinger equation\cite{holland05}here in this paper we explore the question as to whether an ontological model such as the Bohmian one is empirically falsifiable.

The essence of the question as regards empirical equivalence between the standard and the Bohmian model can be seen as follows. Let us focus on the instant \emph{when} an ensemble of particles \emph{begins} to interact with a given potential. If at that \emph{initial instant}, the position probability density of the ensemble is taken to be the \emph{same} in both the standard and any ontological approach, then the time evolved position probability density of the final ensemble as calculated by the equation of motion specified by any ontological model is ensured to be the \emph{same} as that obtained from the standard quantum mechanical formalism. The same equivalence holds good for any other dynamical observable represented by a Hermitian operator. 

The above is the main reason why the ontological models have been dismissed by many as merely of metaphysical interest, having a {}``superfluous
ideological superstructure'' (see, for example, the criticism of the Bohmian model by Heisenberg\cite{heisenberg}. In this paper, we take a fresh look the question of empirical falsifiability of an ontological model like the Bohmian one. The motivation underlying the present study stems from the type of question as was posed by Cushing\cite{cushingbook} : Could it be that the additional interpretive ingredients(such as the notion of particles having objectively defined trajectories) in the Bohmian model might enable testable predictions in a suitable example where the standard version of quantum mechanics has an intrinsic \emph{nonuniqueness} allowing for a number of calculational approaches, but, \emph{prima facie}, none of these can be preferred over the others using a rigorous justification based on the first principles? On this point one may recall that John Bell had once remarked that the Bohmian model of quantum mechanics is experimentally equivalent to the standard version "insofar as the latter is unambiguous"\cite{bell}.

It is from this perspective that the kind of \emph{nonuniqueness} inherent in the standard quantum mechanical calculation of time distributions seems to be a pertinent tool for exploring the possibility of subjecting an ontological model such as the Bohmian one to an empirical scrutiny. 

Time plays a peculiar role within the formalism of quantum mechanics - it differs fundamentally from all other dynamical quantities like position or momentum since it appears in the Schrödinger equation as a parameter, not as an operator.  If the wave function $\psi(x,t)$ is the solution of Schrödinger's equation, then $\psi(x,t_{1})$,$\psi(x,t_{2})$, $\psi(x,t_{3})$. . . determine position probability distributions at the respective different instants $t_{1}$, $t_{2}$, $t_{3}$,. . .  for a fixed region of space, say, between $x$ and $x+dx$. Now, if we fix the positions at$X_{1}$, $X_{2}$, $X_{3}$,. . ., the question arises to whether the quantities $\psi(X_{1},t)$, $\psi(X_{2},t)$, $\psi(X_{3},t)$,. . . can specify the time probability distributions  at respective various positions $X_{1}$, $X_{2}$, $X_{3}$,. . .? However, one can easily see that, although $\int_{-\infty}^{+\infty}|\psi(x,t_{i})|^{2}dx=1$, one would have in general,$\int_{-\infty}^{+\infty}|\psi(X_{i},t)|^{2}dt\neq1$ . Hence, in order to quantum mechanically calculate the time probability distribution, unlike the position probability distribution, we do not readily have an unambiguously defined procedure.

The fundamental difficulty in constructing a self-adjoint time operator within the formalism of quantum mechanics was first pointed out by Pauli\cite{paulitime}. Another proof of the nonexistence of time operator, specifically for the time-of-arrival operator, was given by Allcock \cite{allcock}. Nevertheless, there were subsequent attempts to construct a suitable time operator. For instance, Grot et al. \cite{grot} and Delgado \emph{et al.} \cite{delgado}suggested a time-of-arrival operator for a free particle, and showed how the time probability distribution can be calculated using it; interestingly, such an operator has an orthogonal basis of eigenstates, although the operator is, in general, \emph{not} self-adjoint. 

In recent years there has been an upsurge of interest in analyzing the concept of arrival time distribution in quantum mechanics, for useful reviews on this subject, see \cite{muga,muga1}. Here we note that a number of schemes have been formulated\cite{brouard,yamada,muga95,mckinnon95,delgado99,ali03,baute1,baute2,ah,muga3,pan06,halliwell,leavens93,leavens98,ali1,dambo,hager,kijowski,feynmann, povm,aharonov} for calculating what has been called the \emph{arrival time distribution} in quantum mechanics, for example, the probability current density approach, using the path integral approach, the consistent history scheme, and by using the Bohmian trajectory model in quantum mechanics. However, since there is an inherent ambiguity within the standard formalism of quantum mechanics as regards  calculating such a probability distribution, it remains an open question as to what extent these different approaches can be \emph{empirically tested}. 

There have been several specific toy models that have been to investigate the feasibility of how actually the measurement of a transit time distribution can be performed in a way consistent with the basic principles of quantum mechanics. The earliest proposal for a model quantum clock in order to measure the time of flight of quantum particles was suggested by Salecker and Wigner\cite{salecker}, and later elaborated by Peres\cite{peres}.In effect, this model of quantum clock measures the change in the phase of a wave function over the duration to be measured. Such a model of quantum clock\cite{peres} can be used to calculate the expectation value for the transit time distribution of quantum particles passing through a given region of space. On the other hand, Azebel\cite{azebel1} has analyzed a process in which the thermal activation rate can serve as a clock. Applications of quantum clock models have also been studied for the motion of  quantum particles in a uniform gravitational field by Davies \cite{davies1}and others\cite{ali1}. 

Against the backdrop of such studies, in the present article we proceed as follows. Let us first consider the following simple experimental arrangement. A particle moves in one dimension along the $x$ - axis and a detector is placed at the position $x=X$. Let $T$ be the time at which the particle is detected, which we denote as the \emph{time of arrival} of the particle at $X$. Can we predict $T$ from the knowledge of the state of the particle at the prescribed initial instant?

In classical mechanics, the time of arrival $T$ at $x=X$ of an individual particle with the initial position $x_{0}$ and momentum $p_{0}$  is $T=t(t,x_{0},p_{0})$ which is fixed by the solution of the equation of motion of the particle concerned. But in quantum mechanics, this problem becomes nontrivial in terms of the probability distribution of times, say, $\Pi(T)$ over which the particles are registered at the detector location, say $X$, within the time interval, say, $t$ and $t+dt$ so that $\int_{T_{1}}^{T_{2}}\Pi(T)dT$ is the probability that the particles are detected at $x=X$ between the instants $T_{1}$ and $T_{2}$. In other words, the relevant key question in quantum mechanics is how to calculate $\Pi(T)dT$ at a specified position, given the wave function at the initial instant from which the propagation is considered.

An effort along this direction was made\cite{dambo} by considering the measurement of arrival time using the emission of a first photon from a two-level system moving into a laser-illuminated region.  The probability for this emission of the first photon was calculated for this purpose by specifically using the quantum jump approach. Subsequently, further work\cite{hager}was done on this proposal using Kijowski's distribution\cite{kijowski}.

In this work we address this question of empirical verifiability from a new perspective so that starting from any axiomatically defined \emph{transit/arrival time distribution} one can \emph{directly} relate it to the actually testable results. For this, we first need to evaluate the transit  time distribution $\Pi(t)$ by using one of the suggested approaches. Using such a calculated this $\Pi(t)$, one can then derive a distribution of spin orientations $\Pi(\phi)$ along different directions for the spin-1/2 neutral particles emerging from a spin-rotator (SR)(which contains a constant magnetic field), $\phi$ being the angle by which the spin orientation of a spin-1/2 neutral particle (say, a neutron) is rotated from its initial spin polarised direction. Note that this angle $\phi  $ is determined by the transit time $ (t) $within the SR. Thus, in our scheme, the SR serves as a \emph{{}``quantum clock''}where the basic quantity is  $\Pi(\phi)$which determines the actually observable results which corresponds to the probability distribution of spin orientations along different directions for the spin-1/2 neutral particles emerging from the SR. Such a calculated spin distribution function can be empirically tested by suitably using a Stern-Gerlach(SG)device, as explained later, thereby providing a test of $\Pi(t)$ based on which $\Pi(\phi)$ is calculated. In this article we particularly focus on analysing the application of the Bohmian model in the context of such an example.

In order to recapitulate the essential aspects of this setup \cite{pan06}, we consider spin-$ 1/2 $ neutral objects, specifically, the spin-polarised neutrons. The initial quantum state is given by $ \Psi\left(x,t=0\right) =\psi \left( x,t=0\right) \otimes \chi \left( t=0\right)  $ where $ \chi \left( t=0\right)  $ is the initial spin state taken to be the \emph{same} for \emph{all} neutrons whose spins are oriented along the $\widehat{x}-axis$. The spatial part is taken to be a propagating Gaussian wave packet(for simplicity, taken to be one dimensional) that  moves along the $+\widehat x$-axis, and passes through a region containing constant magnetic field(${\bf B}$), confined between $ x=0 $ and $ x=d $(Fig. 1). The enclosed magnetic field is along the $+\widehat{z}-$axis, and  is switched on \emph{at the instant}(taken to be $t=0$) \emph{when} the peak of the wave packet is at the entry point $x=0$. 

In passing through this bounded region(henceforth designated as a spin rotator(SR)), the spin variable of a neutron interacts with the constant magnetic field, this interaction being mediated by the neutron's magnetic moment $\mu$ as the coupling constant. The angle $\phi$ by which the direction of the initial spin polarisation is rotated is determined by the time $(\tau)$ over which this interaction takes place. Now, if the spatial wave packet is a superposition of energy eigenstates (plane waves), there will be a distribution($\Pi(\tau)$) of times $\tau$ over which the interaction between neutron spins and the enclosed magnetic field takes place, corresponding to the relevant arrival/transit time distribution . Consequently, using the probability distribution function $\Pi(\tau)$, and an appropriate relation connecting the quantities $\phi$ and $\tau$,  one can calculate a distribution of spin orientations  $(\Pi(\phi))$ for the neutrons emerging from the SR. It is, therefore, evident that for calculating $\Pi(\phi)$, the quantum mechanical evaluation of $\Pi(\tau)$ for the neutrons passing through the SR is a key issue. 

We will first elaborate on the relevant setup by discussing of how the determination of the quantity $\Pi \left( \phi \right)$ can actually be tested by using a SG device. Subsequently, we will discuss the key ingredients of the Bohmian procedure that can be adopted in the context of this setup for calculating $\Pi \left( \phi \right)$ in terms of $\Pi \left( t\right)$. The subtleties involved in such a procedure will be critically analysed and directions for further studies will be indicated that are required to explore fully the potentiality of this example for subjecting the Bohmian model to empirical scrutiny.

\section{The setup}

We consider an ensemble of spin $ 1/2 $ neutral particles, say, neutrons having magnetic moment $ \mu  $. The spatial part of the total wave function is represented by a localised narrow Gaussian wave packet $ \psi \left( x,t=0\right)  $ (for simplicity, it is considered to be one dimensional) which is taken to be peaked at $ x=0 $ at tha initial instant $ t=0 $ and moves with the group velocity $ u $. Thus the initial total wave function is given by $ \Psi =\psi \left( x,t=0\right) \otimes \chi \left( t=0\right)  $ where $ \chi \left( t=0\right)  $ is the initial spin state which is taken to be \emph{same} for all members of the ensemble oriented along the $\widehat{x}-axis$.
                                                                                       
The SR used in our setup (Fig. 1) has within it a constant magnetic field $ {\bf B}=B\widehat{\bf z} $ directed along the $ +\widehat{\bf z} $ - axis, confined between $ x=0 $ and $ x=d $. We consider a specific situation where the magnetic field is turned on \emph{while} the peak of the initial wave packet \emph{reaches} at the position $x=0$ at $t=0$. 

Application of the Larmor precession of spin in a magnetic field has earlier been discussed, for example, in the context of the scattering of a plane wave from a potential barrier \cite{baz,buttiker83}. On the other hand, the scheme proposed here explores an application of Larmor precession such that one can empirically test \emph{any} given quantum mechanical formulation for calculating the \emph{arrival/transit time distribution} using Gaussian wave packet.

\begin{figure}[t]
{\rotatebox{0}{\resizebox{10.0cm}{5.0cm}{\includegraphics{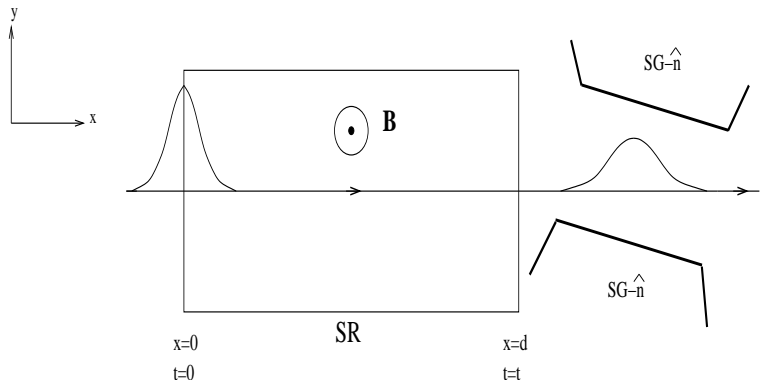}}}}
\caption{\label{fig.1} {\footnotesize Spin-1/2 particles, say, neutrons
with initial spin orientations polarised along the $ +\widehat{\bf x} $
- axis and associated with a localized Gaussian wave packet (peaked
at $ x=0 $, $ t=0 $) pass through a spin-rotator (SR) containing
a constant magnetic field $ {\bf B} $ directed along the $ +\widehat{\bf z} $
- axis. The particles emerging from the SR have a distribution of
their spins oriented along different directions. Calculation of this
distribution function is experimentally tested by measuring the spin
observable along a direction $ \widehat{n}\left( \theta \right) $
in the xy-plane making an angle $ \theta  $ with the initial spin polarised along
$ +\widehat{\bf x} $- axis. This is done by suitably orienting the direction
$ \widehat{n}\left( \theta \right)  $ of the inhomogeneous magnetic
field in the Stern-Gerlach $ \left( SG-\widehat{n}\right)  $ device.}}
\end{figure}

Now, for testing the scheme for calculating the probability density function $ \Pi \left( \phi \right) $ of spin orientations of particles emerging from the SR, we consider the measurement of a spin variable, say $ \widehat{\sigma }_{\theta } $, by a SG device (Fig.1) in which the inhomogeneous magnetic field is oriented along a direction $ \widehat{n}\left( \theta \right)  $ in the xy-plane making an angle $ \theta  $ with the initial spin-polarised direction ($ +\widehat{\bf x} $ axis) of the particles.The initial x-polarised spin state can be written in terms of the z-bases $|\uparrow {\rangle}_z$ and $|\downarrow {\rangle}_z$ as $\chi (0)=1/{\sqrt{2}}~ \left(|\uparrow {\rangle}_z +|\downarrow {\rangle}_z \right)$. While passing through the SR, the spin polarised state rotates only in the xy-plane by an angle $\phi$ with respect to the initial spin orientation along $\widehat{\bf x}$-axis. Such a rotated spin state in the xy-plane can be typically written as $\chi(\phi)=1/{\sqrt{2}}~\left(|\uparrow {\rangle}_z + e^{i \phi} |\downarrow {\rangle}_z \right)$. If one applies the SG-magnetic field along the direction $ \widehat{n}\left( \theta \right)$, the relevant bases states corresponding to the spin operator ${\hat \sigma_{\theta}}$ are respectively $|\uparrow \rangle_{\theta}=1/{\sqrt{2}}~\left(|\uparrow {\rangle}_z + e^{i \theta}|\downarrow {\rangle}_z \right)$, and $|\downarrow \rangle_{\theta}=1/{\sqrt{2}}~\left(|\uparrow {\rangle}_z - e^{i \theta} |\downarrow {\rangle}_z \right)$. Then for such spin measurements, the probabilities of getting $|\uparrow \rangle_{\theta}$ and $|\downarrow \rangle_{\theta}$ are $p_{+}(\theta)={|}_{\theta}{\langle \uparrow}~|~\chi(\phi)\rangle|^2=cos^2 (\theta-\phi)/2$ and $p_{-}(\theta)=~{|}_{\theta}{\langle \downarrow}~|~\chi(\phi)\rangle|^2=sin^2 (\theta-\phi)/2$ respectively.
                                                                                    
\section{The Testable probabilities}
Since we are considering an ensemble of spin-1/2 neutrons passing through the SR where initially \emph{all} members of this ensemble have their spins polarised along, say, the $ +\widehat{\bf x} $ axis, given the \emph{same} initial spin state, the spins of individual members of the ensemble evolve over \emph{different times} (characterised by $ \Pi \left( t\right) $ , the \emph{distribution} \emph{of transit times} within the SR) of duration of the interaction  with the constant magnetic field within the SR.\\

Thus, for the spins of the particles emerging from the SR polarised along different directions (with the respective probabilities $ \Pi \left( \phi \right)$ of making angles $ \phi  $ with the $ +\widehat{\bf x}$ axis), the probabilities of finding the spin component along $ +\theta  $ direction and that along its opposite direction are respectively given by 
\begin{equation} 
P_{+}\left( \theta \right) =\int_{0}^{2n^{\prime}\pi }\Pi \left( \phi \right) {Cos}^2 \frac{\left( \theta -\phi \right)}{2} d\phi
\end{equation}
\begin{equation}
P_{-}\left( \theta \right) =\int_{0}^{2n^{\prime}\pi }\Pi \left( \phi \right)
{Sin}^2 \frac{\left( \theta -\phi \right)}{2} d\phi
\end{equation}
where $P_{+}\left( \theta \right) +P_{-}\left( \theta \right) =1$, and $n^{\prime}$ is any positive number.\\  

It is these probabilities $ P_{+}\left( \theta \right)  $ and $ P_{-}\left( \theta \right)  $ which constitute the actual \emph{observable quantities} in our scheme which are determined by the distribution of spins $ \Pi \left( \phi \right)  $ of the particles emerging from the SR. The theoritical estimations of these probabilities crucially depend on \emph{how} one calculates the quantity $ \Pi \left( \phi \right)$ whose evaluation, in turn, is contingent on the approach adopted for calculating the relevant time distribution $ \Pi \left( t\right) $ mentioned earlier. Now, the important point to stress here is that the specification of such a time distribution is \emph{not} unique in quantum mechanics. For the setup indicated in Fig.1, $ \Pi \left( t\right)  $ represents the \emph{arrival time distribution} at the exit point $ \left( x=d\right)  $ of the SR, which is \emph{also} the \emph{distribution of transit times}$ \left( t\right)  $ as well as the \emph{distribution of times of interaction with the magnetic field} within the SR. Thus, $\Pi(t)$ determining $\Pi(\phi)$ is the key quantity which needs to be evaluated first. In the following section we focus on discussing the Bohmian procedure for calculating   $\Pi(t)$ leading to $\Pi(\phi)$.

\section{The evaluation of $\Pi\left(t\right)$ and $\Pi(\phi)$ using the Bohmian model}

In the Bohmian model\cite{bohm,bohm1}, each individual particle is assumed to have a definite position, irrespective of any measurement. The pre-measured value of position is revealed when an actual measurement is done. Over an ensemble of particles having the \emph{same}wave function $\psi$ , these ontological positions are distributed according to the probability density $\rho=|\psi|^{2}$ where the wave-function $\psi$ evolves with time according to the Schrödinger equation. For the purpose of the present article, the relevant key postulates of the Bohmian model can be expressed as follows:

i) An individual  `particle' embodying the innate attributes like mass, charge, and the magnetic moment has a \emph{definite location} in space at \emph {any} instant, irrespective of whether its position is measured or not. In particular, the pre-measurement value of position is taken to be the \emph{same} as the value revealed by a measurement of position, whatever be the measurement procedure. However, this feature does \emph{not} hold for any other dynamical variable, such as the momentum or spin.

ii) $|\psi|^{2}$dx  is interpreted as the probability of a particle to be \emph{actually present} within a region of space between x and x+dx, instead of Born's interpretation as the probability of \emph{finding} the particle within the specified region.

iii) Consistent with the Schroedinger equation and the equation of continuity being interpreted as corresponding to \emph{actual flow of particles} with velocity $v(x,t)$, the equation of motion of any individual particle is determined by the following equation 
\begin{equation}
{v(x,t)}=\frac{J(x,t)}{\rho(x,t)}=\frac{1}{m}\frac{\partial S(x,t)}{\partial t}
\end{equation}
Here $J(x,t)=\rho(x,t)v(x,t)$ is interpreted as the probability current density of \emph{actual particle flow}and the wave function is considered in the polar form $\psi=Re^{iS/\hbar}$ with $\rho=R^{2}$ where $R$ and $S$ are the real functions of position and time. In the context of our SR example, coming to the question whether there can be predictions obtained from the Bohmian model which may allow to empirically discriminate it from the standard scheme, we analyse this isue as follows.

Here it should be  noted that the Schroedinger expression for the probability current density is inherently \emph{nonunique}. This is seen from the feature that if one adds any divergence-free or constant term to the  expression for $J(x, t)$, the new expression also satisfies the same equation of continuity derived from the Schroedinger equation. However, it has been discussed by Holland\cite{holland2}, followed by others\cite{ali03,pan06}, that the probability current density derived from the Dirac equation for any spin-1/2 particle is \emph{unique}, and that this uniqueness is preserved in the \emph{non-relativistic limit} while containing a spin-dependent term in addition to the Schroedinger expression for $J(x,t)$. Accordingly, a modified Bohmian equation of motion for the spin-1/2 particle has been proposed \cite{holland2} which is of the form given by
\begin{equation}
\label{spincurrent}
{\bf v}({\bf x},t)=\frac{1}{m}\left(\nabla S(x,t)+\nabla \log \rho(x,t) \times \bf{s}(t)\right) 
\end{equation}
where ${\bf s}(t)=\frac{\hbar}{2}\chi(t)\widehat\sigma\chi^{\dagger}(t)$; $\chi(t)$ is the time evolving spin state and the components of $\widehat\sigma$ are the Pauli matrices. The modified Bohmian equation of motion given by Eq.(\ref{spincurrent})reduces to the original Bohmian form given by Eq.(3) if $ \left| \nabla S\right| \gg \frac{\hbar }{2}\left| \nabla \log \rho \right|$, i.e., when the modulus of the spin-dependent term is negligibly small  compared to the spin-independent term. In our setup, the pertinent parameters (viz. the magnitude of the magnetic field, the initial width and the peak velocity of the wave packet) are taken to be such that this condition is satisfied. Hence,  throughout this paper, we will neglect any effect of the spin-dependent term in the modified Bohmian equation of motion. 

Next, we recall the general question of  equivalence between the  standard  version and the Bohmian model of quantum mechanics in predicting the observable results pertaining to any Hermitian operator that has been much discussed\cite{bohm}, along with some controversy \cite{golshani}. Nonetheless, the central point we may stress that, given any example, if and only if the procedure for calculating an observable quantity is \emph{unambiguously defined} in \emph{both} the standard  and the Bohmian versions of quantum mechanics, the very formulation of the Bohmian model  guarantees the equivalence(at-least, in the non-relativistic domain) when the (common) formalism is applied.

Now, coming to our example, for the initial state given by $\Psi\left(x,t=0\right) =\psi \left( x,t=0\right) \otimes \chi \left( t=0\right)$, the time evolution within SR is subject to the interaction Hamiltonian $\vec{\mu\sigma}.B$. Note that as the spins of neutrons interact with the time-independent constant magnetic field, in order to conserve the energy, there will be changes in the momenta of the spatial parts associated with both the up and down spin components -  these changes occur according to the potential energy of the spin-magnetic field interaction that develops corresponding to the up and down spin components respectively. 

In the setup considered here, the relevant parameters(viz. the initial momentum, mass, magnetic moment, and the magnitude of the applied constant magnetic field) are chosen to be such that the magnitude of the potential energy $({vec\mu\sigma} .{\bf B})$ that arises because of spin-magnetic field interaction is exceedingly \emph{small}($\approx 10^{-9}$ times) compared to the initial kinetic energy of the neutrons. This amounts to ensuring the magnitude of the momenta changes within the SR to be negligibly small compared to the initial momentum. Consequently, the time evolutions of the spatial and spin parts can be treated \emph{independent} of each other, with the spatial wave function evolving \emph{freely} within the SR(for  the relevant justification based on a rigorous treatment, see Appendix A ). 

With respect to the  setup shown in Fig.1, the initial wave function is given by
\begin{equation}
\label{inwfn}
\psi(x,t=0) =\frac{1}{\left( 2\pi \sigma ^{2}_{0}\right) ^{1/4}}\exp \left[ -\frac{x^{2}}{4\sigma ^{2}_{0}}+ikx\right] 
\end{equation}
where the wave number $ k=mu/\hbar  $, $ u $ being the group velocity and $\sigma_{0}$ is the initial width of the wave packet. Now, since the spatial wave function is considered to evolve \emph{freely} within the SR, the freely propagating wave function at any instant $t$ is given by 
\vskip -0.9cm
\begin{equation}
\label{tdwfn}
\psi(x, t)=\frac{1}{(2\pi A_{t}^{2})^{1/4}}\exp \left[ -\frac{(x-ut)^{2}}{4\sigma _{0}A_{t}}+ik\left(x-ut/2\right)\right] 
\end{equation}
where $A_{t}=\sigma_{0}\left(1+\frac{i\hbar t}{2m\sigma_{0}^{2}}\right)$.

Writing Eq.(\ref{tdwfn}) in the polar form $\psi(x,t)=R(x,t)e^{iS(x,t)/\hbar}$, and using Eq.(3), the Bohmian trajectory equation in this setup for an $i$-th individual particle having an initial position $x_{i}(0)$ is given by 
\begin{equation}
\label{tra}
x_{i}\left( t\right) = ut +x_{i}(0)\sqrt{1+\beta t^{2}}
\end{equation}
with $ \beta=\frac{\hbar ^{2}}{4m^{2}\sigma ^{4}_{0}}$, and the subscript `$i$' is used as the particle label.

The Bohmian velocity of an \emph{i}-th individual particle can then be written as a function of the initial position by using Eq.(\ref{tra}). The resulting expression is given by
\vskip -0.9cm
\begin{equation}
\label{velindi}
v_{i}(x_{0i},t)=\frac{dx_{i}\left( t\right) }{dt}=u+\frac{x_{0i}\beta t}{\sqrt{1+ \beta t^{2}}}
\end{equation} 
From the  above equation it follows that only those particles initially located around the peak position of the Gaussian wave packet at $ x_{0i}=0 $ follow the Newtonian equation of motion. On the other hand, the particles initially located within the front-half( $ x_{0i}=+ve $) are \emph{ all}  accelerated, while the particles initially lying within  the rear-half( $ x_{0i}=-ve $) are \emph{ all}  decelerated\cite{hollandpage}. It can then be seen that there will be be turning points of the Bohmian trajectories provided the following relation is satisfied between the instant $t^{\prime}$ of the turning point and the Bohmian initial velocity $u$, given by   
\vskip -1.2cm
\begin{equation}
\label{invel}
u=\frac{x_{0i}\beta t^{\prime}}{\sqrt{1+\beta (t^{\prime})^{2}}}
\end{equation}
where $ x_{0i} $ is essentially $-ve$. For the purpose of discussions in this article, we consider the relevant parameters ($u$ and $ \sigma_{0}$) to be such that the above condition is not satisfied, thereby ensuring that, in our example, the particles do \emph{not} have any turning point\cite{leavensturn}. Next, in order to proceed with the Bohmian calculation of $\Pi(\phi)$, we focus on the following two quantities pertaining to neutrons which are initially localised, say, within an arbitrarily chosen region between $x_{i}(0)$ and $x_{i}(0)+dx_{i}(0)$.

First, we note that the probability of  neutrons to be actually localised between $x_{i}(0)$ and $x_{i}(0)+dx_{i}(0)$ is given by $|\psi(x_{i}(0),t=0)|^{2}dx_{i}(0)$. Secondly, we consider the probability of \emph{these} neutrons to cross the region of the  magnetic field enclosed in the SR within the time-interval, say, $T$ and $T+dT$; this is denoted by $\Pi_{B}(T)dT$ where $\Pi_{B}(T)$ is the \emph{arrival time distribution} at the \emph{exit point}$(x=d)$ of the SR, $T$ being the time required by the \emph{i}-th neutron to reach the exit point$(x=d)$ of the SR.

Given the notion of trajectories in the Bohmian model, and taking that \emph{all} the neutrons initially localised within $x_{i}(0)$ and $x_{i}(0) + dx_{i}(0)$ ultimately pass through the entire region of the SR (i.e., all of them cross the exit point$(x=d)$ of the SR, and \emph{none} of them reenters), the above two quantities can be equated so that 
\vskip -0.6cm  
\begin{equation}
\label{bmdis1}
\Pi_{B}(T)dT=|\psi(x_{i}(0),t=0)|^{2}dx_{i}(0)
\end{equation}
\vskip -0.6cm
whence
\begin{equation}
\label{bmdis}
\Pi_{B}(T)=|\psi(x_{i}(0),t=0)|^{2}\frac{dx_{i}(0)}{dT}
\end{equation}
Note that Eq.(11) is a very general relation that determines the arrival time distribution in any causal trajectory model, valid for any form of the initial wave packet. In the particular case of a freely evolving Gaussian wave packet we are considering here in the context of the Bohmian model, we proceed as follows:

By using Eq. (\ref{tra}),  and replacing $t$ by $T$, the quantity $\frac{dx_{i}(0)}{dT}$ on the right hand side of Eq.(11) is calculated by writing $x_{i}(0)$ in terms of $x_{i}(T)$. Further, by invoking the earlier outlined interpretive ingredients $(b)$ and $(d)$ of the Bohmian model, we write the quantity $|\psi(x_{i}(0),t=0)|^{2}$ in terms of $x_{i}(T)$. This is done by using Eq.(3) for $\psi(x, t=0)$, replacing the argument `$x$' of this function by $`x_{i}(0)'$ for varying `$i$' corresponding to different initial positions of the neutrons. Then, putting $x_{i}(T)=d$, using both Eq.(\ref{tra}) and the Schroedinger expression for the probability current density $J(x,t)$, it can be verified that Eq.(\ref{bmdis}) reduces to 
\begin{equation}
\label{equiv}
\Pi_{B}(T)=J(d,T)
\end{equation}
where 
\begin{eqnarray}
\label{curr}
J(d, T)=\frac{1}{\left( 2\pi \sigma _{T}^{2}\right) ^{1/2}}\exp \left\{ - \frac{\left( d-uT\right) ^{2}}{2\sigma _{T}^{2}}\right\}\times \left\{ u+\frac{\left( d-uT \right) \hbar ^{2}T }{4m^{2}\sigma ^{4}_{0}+\hbar ^{2}T ^{2}}\right\}
\end{eqnarray}
is the current density of neutrons at the exit point$(x=d)$ of the SR. Here $\sigma_{T}$ is the width of the wave packet at the instant $T$. Note that $J(d, t)$ given by Eq. (\ref{curr}) is essentially a $+ve$ quantity. Thus, Eq.(\ref{equiv}) shows that the Bohmian \emph{arrival time distribution} at $x=d$ is the same as that given by the probability current density(henceforth designated as PCD) approach\cite{leavens93}; i.e., 
\vskip -0.9cm 
\begin{equation}
\label{alleqv}
\Pi_{B}(T)=\Pi_{PCD}(T)=J(d,T)
\end{equation} 

Then, comes an important element in our example, viz. the consideration of the time evolution of the spin state  according to an equation that is obtained by decoupling the position and the spin parts of the Pauli equation (adopting the `approximation' stated earlier whose justification is indicated in Appendix A),given by

\begin{equation}
\label{e13}
\vec{\mu\sigma}.{\bf B} \ \chi=i\hbar\frac{\partial \chi}{\partial t}
\end{equation}

Now, we recall that in this particular example, the initial spin of any neutron is taken to be polarized along the $+\widehat{\bf x}$-axis; i.e., 
\begin{equation}
\label{inspinwfn}
\chi \left( 0\right) =\left| \rightarrow \right\rangle _{x}=\frac{1}{\sqrt{2}}\left[ \left| \uparrow \right\rangle _{z}+\left| \downarrow \right\rangle _{z}\right]
\end{equation}
Then, since the constant magnetic field ${\bf B}=B\widehat{\bf z}$ within the SR, confined between $x=0$ and $x=d$, is switched on at the instant ($t=0$) the peak of the wave packet reaches the entry point ($x=0$)of the SR, the  time evolved neutron spin state $\chi \left( \tau\right)$  under the interaction Hamiltonian $H=\vec{\mu\sigma }.{\bf B}$ is given by

\begin{equation}
\label{tdspinwfn}
\chi \left( \tau\right) =\exp \left( \frac{-iH\tau}{\hbar }\right) \chi \left( 0\right) =\frac{1}{\sqrt{2}}e^{-i\omega \tau}\, \left[ \left| \uparrow \right\rangle _{z}+e^{i2\omega \tau}\left| \downarrow \right\rangle _{z}\right]
\end{equation}
where $\omega =\mu B/\hbar$ and $\tau$ is the spin-magnetic field interaction time; i.e., the time during which the neutron spin state evolves according to the spin-magnetic field interaction term. Eq.(\ref{tdspinwfn}) implies that after a time interval $\tau$, the spin orientation of a neutron is rotated by an angle $\phi$ with respect to its initial spin-polarised direction where 

\begin{equation}
\label{phi}
\phi=2 \omega \tau
\end{equation}
 
Subsequently, one may put $\tau=T$ in Eq.(\ref{phi}), and use Eq.(\ref{alleqv}) to obtain the probability distribution of spin orientations $(\Pi(\phi))$ of the neutrons emerging from the SR. Then, as applied to our example, the equivalence between the Bohmian scheme and the probability current density approach would seem to follow; i.e., 

\begin{equation}
\label{equivalentphi}
\Pi_{B}(\phi)=\Pi_{PCD}(\phi)=J(d,\phi)
\end{equation} 
\section{Subtleties in the Bohmian calculation of $\Pi(t)$ and $\Pi(\phi)$ }
{\large\bf A.} Let us focus on a critical feature that is involved in obtaining the expression for $\Pi_{B}(T)$ given by Eq.(\ref{equiv})  within the Bohmian framework. This requires using the equality given by Eq.(\ref{bmdis1}), taking it to be true for \emph{all} values of the initial positions $x_{i}(0)$ of the neutrons, including even for \emph{those} neutrons initially located \emph{outside} the SR. Consequently, $\Pi_{B}(T)$ given by Eq.(\ref{equiv}) is the \emph{arrival time distribution} at $x=d$(the exit point of the SR) pertaining to the \emph{entire ensemble} of neutrons, including the neutrons initially located outside the SR. 

Then, since writing the probability  distribution $\Pi(\phi)=J(d,\phi)$ from $\Pi_{B}(T)$ of Eq.(\ref{equiv})using Eq.(\ref{phi}) is contingent upon taking $T=\tau$, this  means regarding the \emph{arrival time distribution} at $x=d$ to be essentially the \emph{distribution of interaction times} $\tau$ over which the spin states of the neutrons belonging to the \emph{entire ensemble}  evolve under the interaction Hamiltonian $\vec{\mu\sigma}.{\bf B}$. 
 
Now, if the above \emph{proviso} is to be satisfied, this would imply that, within the Bohmian model, even if a neutron is located \emph{outside} the region of the SR at the instant ($t=0$) when the magnetic field within the SR is switched on, the spin state associated with \emph{that} neutron would have to be regarded as evolving, under the spin-magnetic field  interaction, from the instant $t=0$ itself. However, a fundamental feature of the Bohmian model is that an individual particle such as a neutron is regarded as a \emph{localised entity} in an objective sense, that has  a \emph{definite location} in space at \emph{any} instant, and which embodies the innate properties such as the mass and the magnetic moment. Further, a relevant crucial point is that the spin-magnetic field interaction term $\vec{\mu\sigma}.{\bf B}$ has the \emph{localised} particle attribute, magnetic moment$(\mu)$,  as the coupling constant. It is, therefore, arguable that in applying the Bohmian scheme to our example, the time evolution of the spin state associated with an individual neutron is to be considered subject to the interaction $\vec{\mu\sigma}.{\bf B}$ \emph{only} if the neutron is \emph{inside} the SR within which the magnetic field is confined. Thus, from this point of view, the earlier mentioned \emph{proviso} would be inadmissible for the neutrons which are \emph{initially} located \emph{outside} the SR. In other words, within the Bohmian model, this would mean taking the ontological position variable to be more fundamental than the global wave function in determining the instant from which the individual particle is subjected to the relevant interaction which is mediated through a localized property of the particle (in this case, the magnetic moment).

Consequently, it would not be  legitimate to take the Bohmian arrival time distribution $(\Pi_{B}(T))$ at the exit point $(x=d)$ of the SR to be the distribution of interaction times $\tau$ for \emph{all} the neutrons belonging to the entire ensemble. This would then mean that the Bohmian probability distribution of spin orientations of neutrons emerging from the SR \emph{cannot} be written as $J(d,\phi)$, thereby implying in the context of our setup, an alternative route for calculating  the quantities $\Pi_{B}(t)$ and $\Pi_{B}(\phi)$. 

Such an alternative Bohmian procedure for calculating $\Pi_{B}(t)$ and $\Pi_{B}(\phi)$ would entail the viewpoint that the spin state of an individual neutron which is initially located \emph{outside} the SR would evolve under the spin-magnetic field interaction that acts using the neutron's localised magnetic moment as the coupling constant only \emph{after} that neutron reaches the entry point $(x=0)$ of the SR. Therefore, for such a  neutron, its arrival time at the exit point$(x=d)$ of the SR would \emph{not} be the time over which its spin-magnetic field interaction occurs in passing through the SR. On the other hand, for a neutron initially located \emph{inside} the SR, its spin state would, of course, start to evolve from the instant $t=0$  itself when the magnetic field is switched on.    

Consequently, in this approach for calculating the Bohmian probability distribution $\Pi_{B}(\phi)$, the time over which the spin of a particle evolves under the magnetic field enclosed within the SR is  essentially the particle's  \emph{transit time} within the SR (\textit{not} the arrival time at $x=d$), which is the time taken by it to cross the region between $x=0$ and $x=d$. Thus, in such a procedure, it is the computation of the \emph{transit time distribution} within the SR (in contrast to the arrival time distribution at $x=d$ computed in the earlier discussed Bohmian procedure) that would provide the prediction of the Bohmian distribution $\Pi_{B}(\phi)$. \emph{Prima facie}, it is thus an open question whether the quantitiy $\Pi_{B}^{\prime}(\phi)$ calculated in this alternative way would agree with the result obtained(Eq.(23)) from the Bohmian calculational scheme discussed in the preceding section - a procedure which, in a sense, ascribes more fundamental ontological status to global wave function associated with an individual particle than to its ontological position variable.

{\large\bf B.} An important point to be stressed here is that the calculational scheme discussed in Section V is specifically couched in terms of a Gaussian wave packet. In particular, the crucial expression given by Eq.(12) is obtained from the generic Eqs. (10) and (11) in a way that involves an explicit use of the Gaussian wave packet. Therefore, it is again another open question whether the equivalence between the Bohmian and the probability current density based approach in calculating the quantity $\Pi_{B}(t)$ would hold good if one considers non-Gaussian wave packets. For example, one may consider a non-Gaussian wave function in the momentum space given by 
\begin{equation}
\label{wvfn-k}
\phi(k) =\frac{N}{\left( 2\,\pi\, \sigma ^{2}_{k}\right) ^{1/4}}\exp \left[ -\frac{(k-k_{0})^{2}}{4\,\sigma ^{2}_{k}}\right]\left(1+\alpha \,\sin \left[{k-k_{0}\over \beta\,\sigma_{k}}\right] \right)\,,
\end{equation}
where $ k $ is the wave number and $N=\left(1+{\alpha^2\over2}(1-e^{-{\pi^2\over8}})\right)^{-1}$ is the normalisation constant. The above function contains four real parameters $\sigma_{k}, k_{0},\,\alpha\,$and $\beta$ among which $\sigma_{k}$ is positive. The salient feature of the corresponding wave-packet, {\it i.e.} $|\phi(k)|^2$, is its infinite tail with the probability of finding particles negligibly small outside a bounded region determined by the parameter $\sigma_{k}$, while the wave packet is asymmetric and it  reduces to the Gaussian form upon continuous decrement of $|\alpha|$ to zero. Note that the infinite tail is in contradistinction to other non-Gaussian forms found in the literature  which are generated 
by truncating the Gaussian distribution. The asymmetry of the wave packet due to the sine function entails a difference between the mean and peak values of the wave packet, and deprives $k_{0}$ from appropriating either of these values. Similarly, the parameter  $\sigma_{k}$ does not denote the standard deviation of $|\phi(k)|^2$.

Using Eq.(\ref{wvfn-k}), the initial wave function in position space is just the Fourier transform  of $\phi(k)$ and is given by
\vskip -0.9cm
\begin{equation}
\label{psi-x0}
\psi(x, t=0)\equiv{1\over \sqrt{2\,\pi}}\int_{-\infty}^{\infty}\,\phi(k)e^{i\,k\,x}\,dk = \frac{N}{\left( 2\,\pi\, \sigma ^{2}\right) ^{1/4}}\exp \left[ -\frac{x^{2}}{4\,\sigma ^{2}}+i\,k_{0}\,x\right]\left(1+i\,\alpha\,e^{-{\pi^2\over16}}\sinh\left[{\pi\,x\over4\sigma}\right]\right)
\end{equation}
 where $\sigma=(2\,\sigma_{k})^{-1}$. 
 
It would thus be an interesting exercise to study the arrival/transit time distribution pertaining to the freely evolving wave packet corresponding to Eq.(\ref{wvfn-k}) and use the results obtained to probe the equivalence between the Bohmian and the probability current density based scheme in calculating the empirically testable probabilities relevant to our setup.
\section{Summary and outlook}

 The setup analysed in this article seems to provide a particularly interesting example in the context of the foundations of quantum mechanics since it is capable of empirically discriminating between not only the various suggested standard quantum mechanical schemes for calculating the arrival/transit time distribution, but it may also enable to empirically \emph{verify} or \emph{falsify} the specific \emph{unique predictions} obtained from any \emph{realist trajectory model} of quantum mechanics like that of the Bohmian type. This is irrespective of the question as regards which standard quantum mechanical scheme is the empirically correct one for calculating the observable probabilities in our example. While  comprehensive computations are being pursued along the lines {\bf A} and {\bf B} indicated in Section V, here it may be useful to summarise  the implications of the different possible outcomes of the relevant experimental study of such a setup:

$a)$ If the experimental results turn out to corroborate the prediction obtained from $\Pi_{B}(\phi)(=\Pi_{PCD}(\phi))$ given by Eq.(\ref{equivalentphi}), this would not only validate a particular quantum approach for calculating the arrival/transit time distribution with its justification provided by the Bohmian model, but it would also mean that if one analyses our example within the Bohmian framework, one would have to infer that although the magnetic field is confined \emph{within} the SR, it  would have to affect, from $t=0$, the time evolution of the spin state of even those individual neutrons that are initially \emph{located} \emph{outside} the SR.Such an effect within the Bohmian model is rather curious  essentially because of the assumed objective (i.e., independent of measurement)localisation of an individual particle, and that the neutron's spin-magnetic field interaction is mediated through the  \emph{localised} particle attribute, viz. magnetic moment, as the coupling constant. Note that this type of question does not arise within the standard version of quantum mechanics because of the absence of the notion of objective localisation of an individual particle.

$b)$ On the other hand, if the alternative Bohmian calculational procedure indicated in {\bf A} of Section V yields the probability distribution $\Pi_{B}^{\prime}(\phi)$ that is \emph{different} from $\Pi_{B}(\phi)$, and if the experimental results are in conformity with $\Pi^{\prime}_{B}(\phi)$, then, within the standard framework of quantum mechanics, this would pose a challenge to find out which particular quantum  scheme, in this case, would yield results in agreement with the Bohmian prediction $\Pi^{\prime}_{B}(\phi)$. 

$c)$ Another possibility, apart from that the experimental results may \emph{not} agree with either $\Pi_{B}(\phi)$ or $\Pi^{\prime}_{B}(\phi)$ for the Gaussian wave packet, stems from the line of study indicated in {\bf B} of Section V - i.e., if by using non-Gaussian wave packet, it is found that the Bohmian prediction in our example disagrees with that obtained from the probability current density based quantum approach, this would again require studying which particular quantum mechanical approach for calculating the arrival/transit time distribution, as applied to our example, would lead to results in agreement with the relevant Bohmian prediction using non-Gaussian wave packet. 

Thus, whatever be the experimental outcome, nontrivial questions  would arise. Further, in view of a number of insightful variants\cite{epstein,roy,ghirardi,holland3} of the Bohmian model that have been proposed using the notion of objectively defined trajectories of individual particles, it should be instructive to analyse the example formulated in this paper in terms of such models in order to find out the extent to which an empirical discrimination is possible. Such studies would reinforce the importance of pursuing a comprehensive experimental programme along the direction that has been indicated here. 

\appendix*\section{A}

The Gaussian wave packet considered in the paper can be regarded as made up of plane wave components. The treatment given here pertains to an individual plane wave component. 

Let the initial total wave function of a particle be represented by $\Psi_{i}=\psi_{0}\otimes\chi$, where  $\psi_{0}=A e^{i k x}$ is the spatial part which is taken to be a plane wave with wave number $k$, and  $\chi=(\frac{1}{\sqrt{2}}(\chi_{+z}+ \chi_{-z})$ is the spin state polarized in the $+x$ direction where $\chi_{+z}$ and $\chi_{-z}$ are the eigenstates of $\sigma_{z}$. Now, we consider that the particle passes through a bounded region(called SR) that contains \emph{constant} magnetic field directed along the $+\widehat{z}$-axis. 

The interaction Hamiltonian is $H_{int}=\vec{\mu\sigma}.\textbf{B}$ where $\mu$ is the magnetic moment of the neutron, $\textbf{B}$ is the enclosed magnetic field and $\vec{\sigma}$ is the Pauli spin vector. Then the time evolved total wave function  at $t=\tau$ after the interaction of spins with the uniform magnetic field is given by
\begin{eqnarray}
\nonumber
\Psi\left(\textbf{x},\tau\right) &=& \exp({-\frac{iH\tau}{\hbar}})\Psi(\textbf{x},0)\\
&=&\frac{1}{\sqrt{2}}\left[\psi_{+}(\textbf{x},\tau)\otimes\chi_{+z}+\psi_{-}(\textbf{x}, \tau)\otimes\chi_{-z}\right]
\label{timeevolved}
\end{eqnarray}
where $\psi_{+}\left({\bf x},\tau\right)$ and 
$\psi_{-}\left({\bf x},\tau\right)$
are the two components of the spinor 
$\psi=\left(\begin{array}{c}\begin{array}{c} 
\psi_{+}\\ \psi_{-}\end{array}\end{array}\right)$ which satisfies the 
Pauli equation. We take the enclosed constant magnetic field as ${\bf B}=B \widehat{z}$. The time evolutions of the position dependent  amplitudes($\psi_+$ and $\psi_-$) of the spin components of the wave function are, thus given by 

\begin{equation}
i\hbar\frac{\partial\psi_{+}}{\partial t}=-\frac{\hbar^{2}}{2m}\nabla^{2}\psi_{+}+\mu  B\psi_{+}
\end{equation}
\begin{equation}
i\hbar\frac{\partial\psi_{-}}{\partial t}=-\frac{\hbar^{2}}{2m}\nabla^{2}\psi_{-} - \mu B\psi_{-}
\end{equation}
Eqs.(A.2) and (A.3) imply that when a neutron having spin up interacts with the constant magnetic field within the SR, the associated spatial wave function $(\psi_{+})$ evolves under a \emph{potential barrier} that has been generated due to the spin-magnetic field interaction, while for a neutron having spin down, the associated spatial wave function $(\psi_{-})$ evolves under a \emph{potential well}. 

Here we will specifically consider the situation where $E>|\mu B|$. This is because, even if one uses low energy or ultra-cold neutrons having kinetic energy of the order of $5 \times 10^{-7} eV$, if the potential energy term ($|\mu B|$) has to exceed the kinetic energy term, one will require the magnitude of the magnetic field to be exceedingly high, to be of the order of $10 T$. Magnetic fields of such high intensity are difficult to produce in the usual laboratory conditions, and therefore for all practical purposes, it suffices to treat the case where $E>|\mu B|$.

Now, since $\psi_{-}$ evolves under a potential well confined between $x=0$ and $x=d$ the reflected and transmitted parts are respectively given by
\begin{equation}
\label{psi-r}
\psi^{-}_{R}= A e^{- i k x}\frac{(k^2- k_1^2)(1-e^{2 i k_1 d})}{(k + k_1)^2- (k- k_1)^2 e^{2 i k_1 d} }
\end{equation}

\begin{equation}
\label{psi-t}
\psi^{-}_{T}= A e^{i k x}\frac{4 k k_1 e^{- i k d} e^{i k_1 d}}{(k + k_1)^2- (k- k_1)^2 e^{2 i k_1 d} }
\end{equation}
where $k=\frac{\sqrt{2 m E}}{\hbar}$ , $k_1= \frac{\sqrt{2 m (E-\mu B)}}{\hbar}$ and $d$ is the width of the SR arrangement which contains the uniform magnetic field. 

On the other hand, for $\psi_{+}$ which evolves under a potential barrier, the expressions for the transmitted and the reflected part are written by replacing all the $k_1$'s in  Eqs.(\ref{psi-r}) and (\ref{psi-t}) by $k_2$ where $k_2=\frac{\sqrt{2 m (E+\mu B)}}{\hbar}$
\begin{equation}
\psi^{+}_{R}= A e^ {-i k x}\frac{(k^2- k_2^2)(1-e^{2 i k_2 d})}{(k + k_2)^2- (k- k_2)^2 e^{2 i k_2 d} }
\end{equation}

\begin{equation}
\psi^{+}_{T}= A e^{i k x}\frac{4 k k_2 e^{- i k d} e^{i k_2 d}}{(k + k_2)^2- (k- k_2)^2 e^{2 i k_2 d} }
\end{equation}

Now, we note that the solutions in these cases, given our initial state, consist of a reflected part traveling in the $-\widehat{x}$-direction and a transmitted part of the wave function, that travels in the $+\widehat{x}$-direction. However, the reflected part of the wave function exists \textit{only} to the left of the SR, and the transmitted part exists \textit {only} to the right of the SR. Our objective here is to calculate  the observable distribution of spins of the  particles emerging from the SR. For this, we need to look at \emph{only} the transmitted part of the wave function. Therefore, the final time evolved transmitted state that is relevant to our purpose is of the following form that represents an entangled state involving the spin and the spatial degrees of freedom
\begin{equation}
\label{fitra}
\Psi_{f}=\frac{N}{\sqrt{2}} ( \psi^{+}_{T}\chi_{+z} + \psi^{-}_{T}\chi_{-z})
\end{equation}
where $N$ is the normalized constant that can be written as $N=\int_{v}(\psi^{+}_T|^{2} + |\psi^{-}_T|^{2}) dv$.

It is then seen that,  in the regime $E>|\mu B|$, using Eqs.(A.5)and (A.7), one can rewrite Eq.(A.8) in the following form
\begin{equation}
\label{finalwfn}
\Psi_{f}= \frac{A e^{i k x}}{\sqrt{2}} (C e^{i \phi_{2}}\chi_{+z} + D e^{i \phi_{1}}\chi_{-z})\equiv \psi_{0}\chi(\phi)
\end{equation}

whereby Eq.(\ref{finalwfn}) is the form of the Larmor precession relation which is valid under the condition $E>|\mu B|$  that has been calculated in terms of the explicit time evolved solutions of Eqs.(A.2) and (A.3).

Here 
\begin{equation}
\label{c}
C=\sqrt{Re(\psi^{-}_{T})^{2}+Im(\psi^{-}_{T})^{2}}
\end{equation}
\begin{equation}
\label{d}
D=\sqrt{Re(\psi^{+}_{T})^{2}+Im(\psi^{+}_{T})^{2}}
\end{equation}
\begin{equation}
\label{phi1}
\phi_{1}= tan^{-1}\frac{Im(\psi^{-}_{T})}{Re(\psi^{-}_{T})}
\end{equation}
and
\begin{equation}
\label{phi2}
\phi_{2}= tan^{-1}\frac{Im(\psi^{+}_{T})}{Re(\psi^{+}_{T})}
\end{equation}
Note that, by using Eq.(\ref{psi-t}), one can find that
\begin{eqnarray}
\label{repsi-t}
Re(\psi^{-}_{T})= \frac{8k k_1(k^2+k_1^2)sin(k d)sin(k_1 d)+16k^2 k_1^2cos(k d)cos(k_1 d)}{(k+k_1)^4+(k-k_1)^4-2(k+k_1)^2(k-k_1)^2 cos(2 k_1 d)}
\end{eqnarray}
\begin{eqnarray}
\label{impsi-t}
Im(\psi^{-}_{T})=\frac{8k k_1(k^2+k_1^2)cos(k d)sin(k_1 d)-16k^2 k_1^2sin(k d)cos(k_1 d)}{(k+k_1)^4+(k-k_1)^4-2(k+k_1)^2(k-k_1)^2 cos(2 k_1 d)}
\end{eqnarray}
Similarly, by using Eq.(A.7), one can obtain the expressions for $Re(\psi^{+}_{T})$ and $Im(\psi^{+}_{T})$ given by
\begin{eqnarray}
\label{repsi+t}
Re(\psi^{+}_{T})=\frac{8k k_2(k^2+k_2^2)sin(k a)sin(k_2 d)+16k^2 k_2^2cos(k d)cos(k_2 d)}{(k+k_2)^4+(k-k_2)^4-2(k+k_2)^2(k-k_2)^2 cos(2 k_2 d)}
\end{eqnarray}
\begin{eqnarray}
\label{impsi+t}
Im(\psi^{+}_{T})=\frac{8k k_2(k^2+k_2^2)cos(k a)sin(k_2 d)-16k^2 k_2^2sin(k d)cos(k_2 d)}{(k+k_2)^4+(k-k_2)^4-2(k+k_2)^2(k-k_2)^2 cos(2 k_2 d)}
\end{eqnarray}

Now, let us examine in what limit one can obtain the standard expression for Larmor precession. 

Considering the more stringent limiting condition $E>>|\mu B|$, it can be seen that when the kinetic energy term of the Hamiltonian is much larger than the potential energy term, one can assume that the time evolution of the wave function occurs effectively due to a \emph{very shallow well} and a \emph{very low barrier}. This situation would correspond to the entire wave being transmitted while picking up just a phase.
 
From the expressions for $k, k_1, k_2$ given earlier, in the limit $E>>|\mu B|$, we find that $k\approx k_1\approx k_2$. Then, in order to get the standard expression for Larmor precession, one can first set $k=k_1=k_2$ in Eqs.(\ref{repsi-t}),(\ref{impsi-t}), (\ref{repsi+t}), and in (\ref{impsi+t}), \emph{except} when they appear inside the sine or cosine functions, since the latter terms are much more sensitive to the differences in the values of $k, k_1, k_2$. Eqs. (\ref{repsi-t}) and (\ref{impsi-t}) would then simplify to  
\begin{equation}
Re(\psi^{-}_{T})= sin(k d)sin(k_1 d)+cos(k d)cos(k_1 d)
\end{equation}
\begin{equation}
Im(\psi^{-}_{T})=sin(k_1 d)cos(k d)-cos(k_1 d)sin(k d)
\end{equation}
Using the above expressions in Eqs.(\ref{c}) and (\ref{phi1}), we find that $C=1$ and $\phi_1=(k_1-k)d$. Similarly, rewriting equations (\ref{repsi+t}) and (\ref{repsi+t}), and using Eqs.(\ref{d}) and (\ref{phi2}), we get $D=1$ and $\phi_2=(k_2-k)d$. Therefore, Eq.(A.9) reduces to the form
\begin{equation}
\label{psiwfn1}
\Psi_{f}= \frac{Ae^{i k x}}{\sqrt{2}} \left( e^{i (k_2-k)d}\chi_{+z} + e^{i (k_1-k)d}\chi_{-z}\right)
\end{equation}
Now, remembering that $k=\frac{\sqrt{2 m E}}{\hbar}$ , $k_1= \frac{\sqrt{2 m (E-\mu B)}}{\hbar}$ and $k_2= \frac{\sqrt{2 m (E+\mu B)}}{\hbar}$, we can binomially expand $k_1$ and $k_2$ around $k$, and keep terms up to the order of $\frac{\mu B}{E}$, since we have already stipulated the condition $k\approx k_1\approx k_2$. Then, $(k_1-k)d= - k\frac{\mu B}{2 E}d$, and using $v=\frac{\hbar k}{m}$, we can write $(k_1-k)d= - \frac{\mu B}{\hbar}\frac{d}{v}=  \omega t$. 

Similarly, $(k_2-k)d=  \frac{\mu B}{\hbar}\frac{d}{v}=  -\omega t$. Therefore, we can write Eq.(\ref{psiwfn1}) as
\begin{eqnarray}
\Psi_{f} &=& \frac{A e ^{i k x}}{\sqrt{2}} \left( e^{- i \omega t}\chi_{+z} + e^{i \omega t}\chi_{-z}\right)\\
\nonumber
&&=\psi_{0}\frac{e^{- i\phi/2}}{\sqrt{2}}\left(\chi_{+z}+ e^{i\phi}\chi_{-z}\right)\equiv e^{- i\phi/2}\psi_{0} \chi(\phi)
\end{eqnarray}
where $\phi=2\omega t$. The spin part of this equation is Eq.(17) given in the text, written by considering the time evolutions of the spatial and the spin parts to be \emph{independent} of each other, with the spatial wave function evolving \emph{freely} within the SR.  

It is important to stress here that the rigorous treatment given above of the time evolution of the quantum state of a neutral spin-1/2 particle in a constant magnetic field reveals that the standard expression for Larmor precession given by Eq.(14) in the text is essentially justified in the \emph{limit} where the kinetic energy term is \emph{much higher} than the potential energy term. Thus, subject to this condition being satisfied, the assumption of mutual independence of the evolutions of the spatial and the spin parts of the total wave function that has been used in the text is justified for the setup considered in this article. 
   
\section*{Acknowledgements}
\vskip -0.5cm
We thank Arka Banerjee for helpful interactions. AKP acknowledges useful discussions during his visits to  Perimeter Institute, Canada,  Centre for Quantum Technologies, National University of Singapore, and Atominstitut, Vienna. DH thanks DST, Govt. of India for the relevant project support. DH also thanks the Centre for Science and Consciousness, Kolkata. AKP acknowledges the Research Associateship of Bose Institute, Kolkata. \\

\end{document}